\documentclass[nofootinbib,twocolumn,superscriptaddress,prd]{revtex4-2}

\usepackage{amsfonts} 
\usepackage{graphicx} 
\usepackage{amsmath}
\usepackage{mathtools}
\usepackage{latexsym}
\usepackage{geometry}
\usepackage{array}
\usepackage{epsfig}
\usepackage{epstopdf}
\usepackage{euscript}
\usepackage{tabularx}
\usepackage{bm}
\usepackage{enumitem}
\usepackage{upgreek}
\usepackage{mathrsfs}
\usepackage[utf8]{inputenc}
\usepackage{amssymb}
\usepackage{listings}
\usepackage{xcolor}
\usepackage[normalem]{ulem}
\usepackage[misc]{ifsym}

\begin{document}

\title{ Tutorial on running median subtraction filter with application to searches for exotic field transients in multi-messenger astronomy}

\author{Arko P. Sen}
\email{{Corresponding author email: asen@unr.edu}}
\affiliation{Department of Physics, University of Nevada, Reno,{Nevada},USA} 

\author{Andrey Sarantsev} 
\affiliation{Department of Mathematics and Statistics, University of Nevada, Reno,{Nevada},USA}

\author{Geoffrey Blewitt}

\affiliation{Nevada Bureau of Mines and Geology, University of Nevada, Reno,{Nevada},USA}
\affiliation{Department of Physics, University of Nevada, Reno,{Nevada},USA}


\author{Andrei Derevianko}
\affiliation{Department of Physics, University of Nevada, Reno,{Nevada},USA}


\begin{abstract}

Running Median Subtraction Filter (RMSF) is a robust statistical tool for removing slowly varying baselines in data streams containing transients (short-duration signals) of interest. 
In this work, we explore the RMSF performance and properties using simulated time series and analytical methods. We study the RMSF fidelity in preserving the signal of interest in the data using (i) a Gaussian pulse and (ii) a transient oscillatory signal. Such signals may be generated by hypothetical exotic low-mass fields (ELFs) associated with intense astrophysical events like binary black hole or neutron star mergers. We consider and assess RMSF as a candidate method to extract transient ELF signals. RMSF operates by sliding a window across the data and subtracting the median value within each window from the data points. With a suitable choice of running window size, RMSF effectively filters out baseline variations without compromising the integrity of transients.  The RMSF window width is a critical parameter: it must be wide enough to encompass a short transient  but narrow enough to remove the slowly varying baseline.  
We show that the RMSF removes the mean of a normally distributed white noise while preserving its variance and higher order moments in the limit of large windows. In addition, RMSF does not color the white noise stream, i.e., it does not induce any significant correlation in the filtered data. Ideally, a filter would preserve both the signal of interest and the statistical characteristics of the stochastic component of the data, while removing the background clutter and outliers. We find the RMSF to satisfy these practical criteria for data pre-processing. While we rigorously prove several RMSF properties, the paper is organized as a tutorial with multiple illustrations of RMSF applications.

\end{abstract}

\maketitle 

\section{Introduction} 
\label{Sec:Intro}

The search for predicted transient signals in experimental data is an obvious path to discovery. It is therefore important to assess the performance of methods proposed to detect such transients.  One example of predicted transients is motivated by hypothetical exotic low mass fields that could be emitted by intense astrophysical events such as mergers of black holes or neutron stars{~\cite{Dailey2020,khamis2024}}. As we shall explore, such transients may be characterized as sinusoidal waves within a Gaussian envelope, traveling near the speed of light.

Whatever the form of the signal, a typical search for sufficiently short transients in experimental data benefits from the removal of slowly varying background clutter from the data. In particular, it improves the stationarity\footnote{A data stream is considered stationary when its mean, variance, and auto-correlation structure do not change with time.} of the stochastic component of the data stream. 


There are different types of filters used in practice to remove such backgrounds. The running median subtraction filter (RMSF) is the focus of our {article}. RMSF is a type of non-linear filter~\cite{Barnett1989} which is robust and has an advantage over linear filters in preserving {\it transient}~\footnote{Transient signals refer to short-term fluctuations or patterns within a dataset that are temporary and not indicative of a long-lasting trend.} signal component in the data. A {\it robust} filter is one which is not unduly affected by outliers or deviations from the assumptions made in traditional statistical methods. Outliers are those data values which lie significantly outside the average spread in the data stream. As the median of a data set is not a linear combination of the data values, it makes RMSF a {\it non-linear} filter. The RMSF origins can be traced down to several published works during the 1970s~\cite{Andrews1972, Claerbout1973, Frieden1976, Tukey1977}. We refer the reader to a review~\cite{Ivk2015} on a multitude of RMSF applications. It has been shown that the RMSF is more robust~\cite{Claerbout1973,Frieden1976} and more efficient in removing slowly varying background clutter, or {\it baseline} curvatures in data streams, than a running average subtraction~\cite[Chap. 5, Sec. 5.5.4]{Gregory}. 

\begin{figure}[ht!]
\centering
\includegraphics[width=1.0\columnwidth]{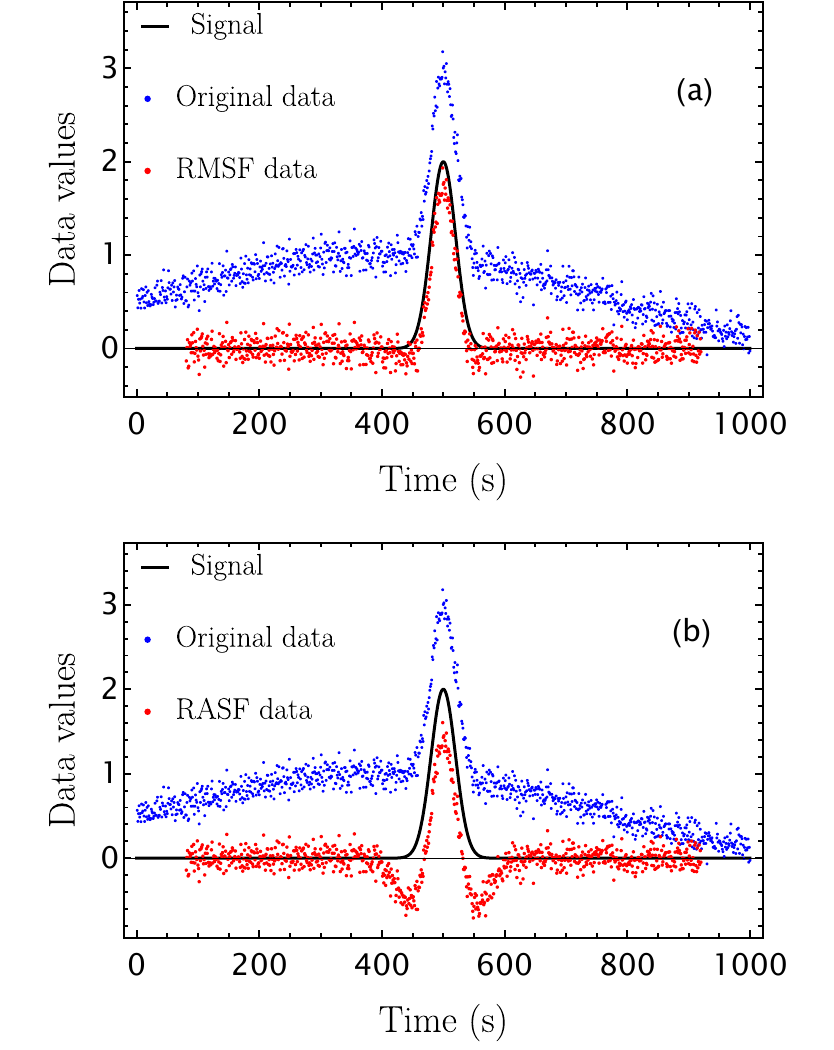} 
\caption{Comparison of a running median subtraction filter (RMSF), panel (a), with a running average subtraction filter (RASF), panel (b). The simulated time series is a sum of a normally distributed white noise, a sinusoidal baseline variation, and a Gaussian signal. The two panels compare the original data stream (in blue) with the transformed data (in red) after applying the filters. It can be seen that RMSF removes the slowly varying sinusoidal background without distorting the Gaussian signal (in black), whereas RASF removes the varying baseline but significantly distorts the signal by introducing two dips on both the sides of the signal, see Sec.~\ref{Sec:Sim}.} \label{Fig:Filter-Plots}
\end{figure}

To motivate our discussion, we generate a time series data stream which is the sum of a normally distributed {\it white noise} ({a data stream of white noise is uncorrelated}), a transient Gaussian signal, and a very slowly varying sinusoidal baseline. The results of applying RMSF to this data stream are shown in Fig.~\ref{Fig:Filter-Plots}(a). Clearly, the RMSF filter effectively removes the slow varying baseline, while preserving the transient signal component. In contrast, a  running {\it average} subtraction filter (RASF) distorts the signal profile by introducing two dips around the signal, see Fig.~\ref{Fig:Filter-Plots}(b). The details of this analysis will be discussed in Sec.~\ref{Sec:Sim}.

This paper is organized as follows: In Sec.~\ref{Sec:Def}, we review the sample mean and sample median; introduce the two types of filters: RMSF (median subtraction) and RASF (average subtraction). In Sec.~\ref{Sec:Def}, we also discuss RMSF effects on monotonic and non-monotonic deterministic functions. The application of RMSF and its comparison with RASF is discussed in Sec.~\ref{Sec:Sim}. The results of applying RMSF to a Gaussian and an exotic field transients are discussed in Sec.~\ref{Sec:Sig-RMSF} and \ref{Sec:Sig-ELF} respectively. Finally, in Sec.~\ref{Sec:Norm-RMSF}, we analyze the RMSF effect  on a normally distributed white noise and draw conclusions in Sec.~\ref{Sec:Conclusions}.

\section{Basic definitions and properties}\label{Sec:Def}
To begin with, we review the fundamental statistical concepts in data analysis, focusing on the sample mean, median, covariance and correlation. Followed by that, we provide definitions for the running median and average subtraction filters and explain the calculation procedure. The window size is a critical factor for these filters. As shown in Fig.~\ref{Fig:Filter-Plots}(a), RMSF can preserve the signal of interest in a data stream by removing the unwanted baseline variation, which only happens if the window width is much larger than the signal duration, see Sec.~\ref{Sec:Sim}. We will also discuss the effect of RMSF on different types of baseline functions such as constant, linear and monotonic (increasing or decreasing) functions. RMSF with any arbitrary window size can eliminate these baselines. However, if a baseline function has a point of extremum in a window, then RMSF fails to eliminate it.

The sample mean (average) and median are two well-known statistics\footnote{A {\it statistic} is a function of the random variables in a data set that summarizes or represents a specific characteristic of the dataset~\cite[Chap. 6, Sec. 6.1]{Gregory}. It is computed from the sample data and used to infer properties about the population from which the sample is drawn.} in data analysis; we employ these statistics in defining the running median and average subtraction filters. Given a discrete data set, $\{x_j\}\equiv x_1,x_2,...,x_N$, the sample {\it mean} or {\it average}, $\bar{x}$, is the value $m$ for which the sum of all $\left(x_j-m\right)^2$ is minimized~\cite{Claerbout1973}\cite[Chap. 1, Example 1.1(i)]{Lehmann}, i.e.,
\begin{align}
   &\nonumber \frac{\partial}{\partial m}\sum_{j=1}^N\left(x_j-m\right)^2\Biggr\rvert_{m=\bar{x}}=0\,,\\
   \Rightarrow~& \bar{x}=\frac{1}{N}\sum_{j=1}^N x_j.\label{Eq:Mean}
\end{align}
In contrast, the {\it median} of the data array $\{x_j\}$ is the value $m$ for which the sum of all~$|x_j-m|$ is minimized \cite{Claerbout1973}, i.e.,
\begin{align}
    &\nonumber \frac{\partial}{\partial m}\sum_{j=1}^N |x_j-m|\Biggr\rvert_{m=\text{median}}=0\,,\\
  \nonumber \Rightarrow~& \sum_{j=1}^N \text{sgn}\left(x_j-\text{median}\right)=0\,,  
\end{align}\label{Eq:sgn}
where $\mathrm{sgn}(z)$ is a sign function which is $+1$ if $z>0$, $-1$ if $z<0$, and it is 0 if $z=0$. 
Next, we sort the data stream $\{x_j\}$ in ascending order and denote the sorted array as $\{x_{(j)}\}$, so that $x_{(1)}\le x_{(2)}\le...\le x_{(N)}$.
Then, if $N$ is odd, the median is the central value of the sorted stream $\{x_{(j)}\}$, or if $N$ is even, it is the average of the two central points:
\begin{align}
\begin{split}
    & \mathrm{Median}\left[\{x_{j}\}_{j=1}^N\right]\\ & =\left\{
\begin{array}
[c]{cl}%
x_{\left(\frac{N+1}{2}\right)}  \,, &  \text{if $N$ is odd} \\
\frac{1}{2}\left[x_{\left(\frac{N}{2}\right)}+ x_{\left(\frac{N}{2}+1\right)}\right] \,, &  \text{if $N$ is even.}
\end{array}
\right.
\end{split}
\label{Eq:Median}
\end{align} 
Notice that the result does not depend on the sorting being either in ascending or descending order.
 Our notation for the $\{x_j\}$ array median, $\mathrm{Median}\left[\{x_{j}\}_{j=1}^N\right]$, specifies the first and the last point of the array. This notation becomes useful when we consider medians over running windows, i.e., a median over a continuous sub-array of $\{x_j\}$. For example, if the window starts at $j=50$ and ends at 100, we would explicitly indicate the window boundaries: $\mathrm{Median}\left[\{x_j\}_{j=50}^{100}\right]$. 

Some additional insights can be gained in the limiting case of a continuous random variable~$X$. In this case, the mean is the expectation value of $X$,
\begin{align}
    \left\langle X \right\rangle=\int_{-\infty}^{\infty} x ~p(x) dx\,,
\end{align}
where $p(x)$ is the probability density function. Note that $x$ represents realizations of the random variable~$X$. The median of $X$ is the value which has half the area under the probability density function lying to one side of it, and the other half lying on the other side,
\begin{align}
\begin{split}
 &\text{Prob}\left(X\leq\text{median}\right)\\ & =\text{Prob}\left(X\geq\text{median}\right)=\frac{1}{2}\,,    
\end{split}
 \end{align}
where Prob$\left(a\leq X\leq b\right)=\int_a^bp(x)dx$ is the probability of finding $X$ in the interval $a\leq X\leq b$. For symmetric distributions, such as the normal distributions, $\text{median} = \left\langle X \right\rangle$.

Suppose $\{x_j\}$ is an independent and identically distributed (IID) data stream with an odd number sample size of $N$. The median of this data stream is well defined and is equal to one of the data points that occupies the central position in a sorted stream, see~\eqref{Eq:Median}. Hence, the probability of any of the data points being the median is
\begin{align}
    \mathrm{Prob}\left(x_j=\mathrm{Median}\left[\{x_{j}\}_{j=1}^N\right]\right)=\frac{1}{N} \label{Eq:Prob}.
\end{align}
Indeed this is true by symmetry, as swapping any of these data points does not modify the probability and the distribution of the data stream. Now we remind the reader the conventional definition of the variance of a random variable $X$,
\begin{align}
  &  \sigma^2_X= \langle (X - \langle X \rangle)^2\rangle = \langle X^2\rangle - \langle X\rangle^2\,,
\end{align}
with $\sigma_X$ being the standard deviation of the random variable. The covariance and correlation of two random variables $X$ and $Y$ is defined as:
\begin{align}
\begin{split}
\mathrm{Cov}(X, Y) &= \langle(X - \langle X\rangle)(Y - \langle Y\rangle)\rangle \\ & = \langle XY\rangle - \langle X\rangle \langle Y\rangle,\\
\mathrm{Corr}(X, Y) & = \frac{\mathrm{Cov}(X, Y)}{\sigma_X \sigma_Y}\,,    
\end{split}
\end{align}
where $-1\leq \mathrm{Corr}(X, Y) \leq 1 $ by {\it the Cauchy-Schwarz inequality}~\cite[Chap. 4, Example 4.7.4]{Casella}:
\begin{equation}
    \label{eq:Cauchy}
    |\mathrm{Cov}(X, Y)| \le \sigma_X\sigma_Y.
\end{equation}
Note that, unlike the covariance, the correlation $\mathrm{Corr}(X, Y) $ is invariant over scaling (i.e. replacing $X$ with $aX$ or $Y$ with $bY$, with constants $a, b > 0$).

The next ingredient in defining the filter is the definition of the running window. A window contains a continuous sub-array of our data stream. The window size $N_w$ remains constant during the filter application. The filter generates a single processed data point, based on $N_w$ original data points contained in the window. We slide the window center incrementally by one point. This process generates a sequence of windows and a stream of processed data. Notice that unless $N_w=1$, these windows overlap.

For sake of simplicity, we choose the window size $N_w$ to be an {\it odd} number, imposing equal number of data points,
\begin{align}
 M=\frac{N_w-1}{2}\,,\label{Eq:M}
\end{align}
relative to the midpoint of the window. {The number of points in the window, $N_w$, should be chosen such that the window  is much wider than the signal width, but narrower than the characteristic time of baseline variation~\cite[Chap. 5, Sec. 5.5.4]{Gregory}, see Sec.~\ref{Sec:Sim}.}

A window centered on index $k$ can be defined as the sub-array of data $\{x_{k-M},...,x_k,...,x_{k+M}\}=\{x_j\}_{j=k-M}^{k+M}$. The midpoint $k$ of this window is the index of the filter output data $\tilde{x}_k$,
\begin{align}
   \Tilde{x}_k=x_k-\mathrm{Median}\bigl[\{x_j\}_{j=k-M}^{k+M}\bigr].\label{Eq:med-sub}
\end{align}
When sliding the window, the first window midpoint is located at $k=M+1$ and that of the last window is at $k=N-M$ of the original data stream. Therefore, the length of the filtered data array $\{\tilde{x}_k\}$ is reduced to $N-2M$ points. 

Let us illustrate the RMSF calculation with a data stream $\{x_j\}=\{0.21,0.52,0.65,0.15,0.72\}$ with $N=5$ elements. Based on~\eqref{Eq:Median}, the median of this data stream is 0.52. We use a running window of size $N_w=3$. {Note that,} $N_w=3$ corresponds to $M=1$ points on the left or right side of the window, see~\eqref{Eq:M}. The first running window is centered on the second element of the data and contains $\{0.21,0.52,0.65 \}$ sub-array. The median is 0.52 and the RMSF-processing~\eqref{Eq:med-sub} results in $\tilde{x}_1=0.52-0.52=0$. Notice that we skipped $j=1$ in the indexing of the filtered data, as a consequence of~\eqref{Eq:med-sub}. The second window is obtained by sliding the first window to the right by one element, i.e., this window is centered on the third element of $\{x_j\}$. The median of $\{0.52,0.65,0.15 \}$ is 0.52 and the RMSF yields $\tilde{x}_2=0.65-0.52=0.13$. Finally, the median of the third window, centered on the fourth element of $\{x_j\}$, is 0.65. Thereby, $\tilde{x}_3=0.15-0.65=-0.5$. The resulting RMSF-processed data array is $\{\varnothing,0,0.13,-0.5,\varnothing\}$, where the symbol $\varnothing$ denotes the data which were not assigned. Notice that the original data stream with $N=5$ elements has been reduced to $N-2M=3$ values.

We also define the running average subtraction filter (RASF), which transforms the data set $\{x_j\}$ as
\begin{align}
    x'_k=x_k-\frac{1}{N_w}\sum_{j=k-M}^{k+M}x_j\label{Eq:mean-sub}\,,
\end{align}
i.e., an average over the running window is subtracted from the original data at the window center. Using the same data set as in the previous paragraph, the RASF processed data is $\{\varnothing, 0.06,0.21,-0.36 ,\varnothing\}$. The sequence of running windows is identical to the RMSF procedure, but here we take the {\it average} value (instead of the median) over each window and subtract it from the original data value at the window center. Similar to the RMSF, the number of elements has been reduced in the RASF-processed data.

Next, we  prove that RMSF with an arbitrary  odd-sized window eliminates 
a constant, a linear, and a monotonically increasing/decreasing functions $f(t)$ (without any stochastic components). Suppose the data stream is constant, $x_j=c$. The median over any window (independent of size and position) is $c$. This implies, see~\eqref{Eq:med-sub}, the filtered result of 
$    \tilde{x}_k=c- \mathrm{Median}\left[\{x_j\}\right]=c-c=0.
$
Therefore, any constant function is removed by RMSF with the running window  of arbitrary width.

Now take a linear function $f(t_j)=a\, t_j=x_j$, to be the data array, sampled on a time grid. Notice that to compute the median, we need to sort the original data array either in the ascending or descending order, with the median  being the central value in such an ordered array. 
The data array within any arbitrarily-sized window is already sorted (either in descending or ascending order depending on the sign of $a$). Then the median is  
$a\,t_k$, where $k$ indexes the window midpoint. Thus, the RMSF processed result vanishes, as
$\tilde{x}_k=x_k- \mathrm{Median}\left[\{x_j\}\right]=at_k-at_k=0.$

\begin{figure}[ht!]
    \centering  \includegraphics[width=0.95\columnwidth]{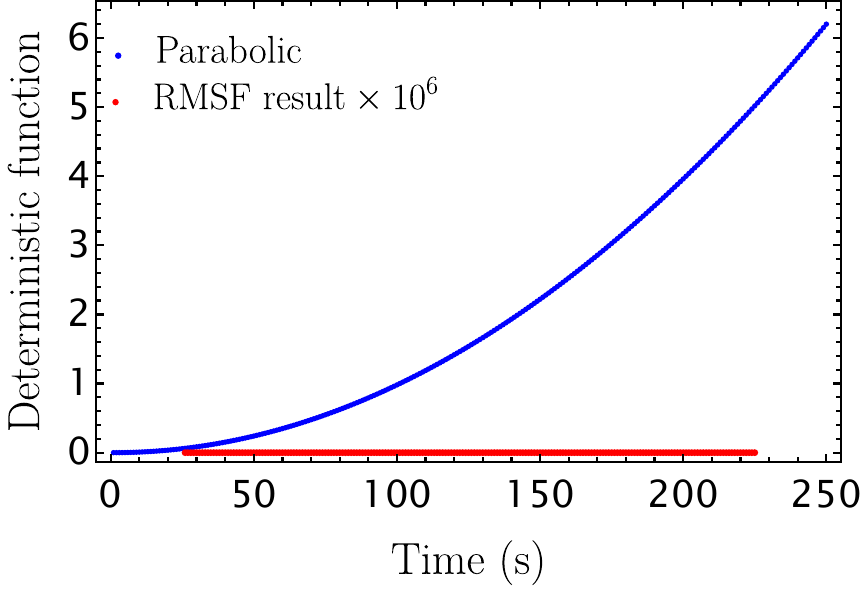}
    \caption{Application of the running median subtraction filter (RMSF) with a window size $N_w=51$ to a parabolic function, $0.0001\, t_j^2$, {with} $t_j=j\Delta_t \,\mathrm{seconds}$. Here, the time step $\Delta_t=1$~s and $j=\overline{1,250}$. The RMSF output (red dots) has been multiplied by a factor of $10^6$ to amplify any small non-zero values. We see that the parabolic function is completely eliminated.}\label{Fig:PB}
    \end{figure}

The extension of these properties to the more general monotonically increasing/decreasing functions $f(t_j)$ is straightforward, as the original data arrays $x_j=f(t_j)$ come already sorted; thereby the median equals to $f(t_k)$ leading to vanishing filtered values, $\tilde{x}_k=0$. This useful property is illustrated in Fig.~\ref{Fig:PB} for a monotonically increasing parabolic function. 

\begin{figure}[ht!]
\centering
\includegraphics[width=1.05\columnwidth]{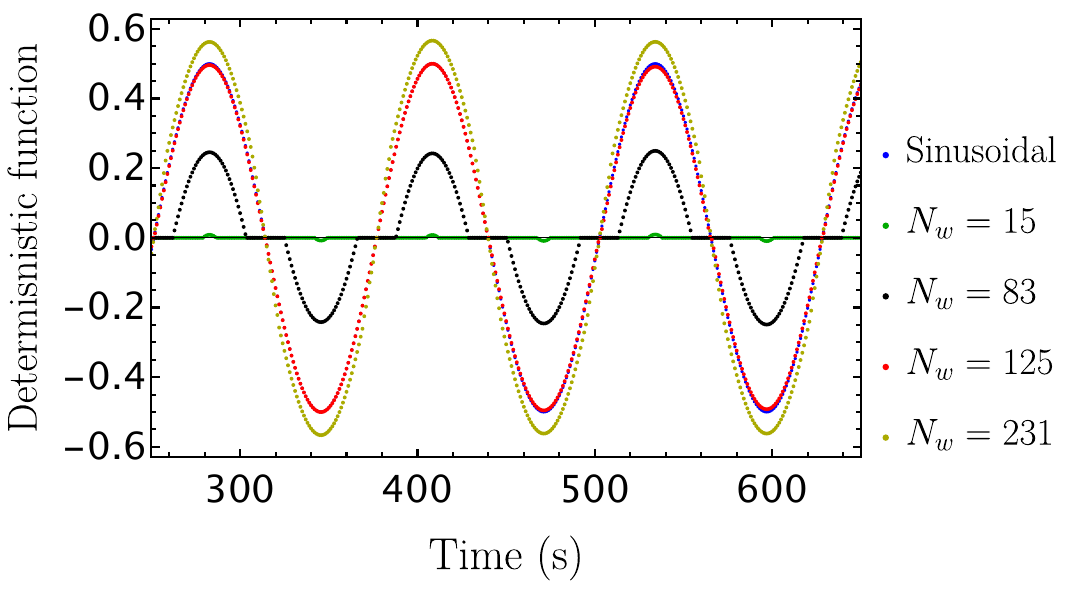} 
\caption{Effect of the running median subtraction filter (RMSF) on a sinusoidal function, with an oscillation period $T_o\approx126$~s. RMSF preserves the function in the limit $N_w\Delta_t\approx qT_o$ where $q$ is an integer and eliminates it when $N_w\Delta_t \ll T_o/2$. Here, RMSF with $N_w=125$ is approximately comparable to integer multiple of $T_o$ and hence it preserves the function. When the median in a window is less than zero, then the filtered function can have higher magnitude than the original, as shown in the case of $N_w=231$. On the other hand, when the median in a window is positive and less than the function value, the filtered data is dampened, as shown in the case for $N_w=83$. RMSF with $N_w=15$ eliminates the function except at the points of extrema.
In this simulation we considered a uniform grid with $N=1000$ points and a time step $\Delta_t=1$~s. We show the results from 250~s to 650~s here.} \label{Fig:Sine-Baseline}
\end{figure}

An important corollary  is that if the function $f(t_j)$ is {\em non-monotonic}, i.e.  within a certain window it  reaches an extremum, then the filter cannot fully eliminate the function within that window. As an example, Fig.~\ref{Fig:Sine-Baseline} shows the RMSF effect on a sinusoidal  function $f(t_j)=0.5\sin\left(2\pi t_j/T_o\right)$ with different window sizes. 
When the window width $N_w\Delta_t$ is comparable to integer multiples of $T_o$, the filter preserves the function. However, when $N_w\Delta_t \ll T_o/2$, the filter  eliminates the function except in those windows that contain the function extrema.

In the case of oscillating functions, RMSF can amplify the original function when, in a particular window, $\mathrm{Median}\left[\{f(t_j)\}_{j=k-M}^{k+M}\right]<0$. This can happen when in a window there are more negative values of the function than the positive values. Hence, in such a case, applying Eq.~\eqref{Eq:med-sub} to the function $f(t_j)$, we get $\tilde{f}(t_k)>f(t_k)$, i.e. the RMSF-processed result exceeds the original values resulting in the increase in the amplitude of the filtered function. This happens near the extrema of the function and is observed to occur in the ranges $1.5T_o \lesssim N_w\Delta_t<2T_o$, $3.5T_o \lesssim N_w\Delta_t<4T_o$, and so on. However, when $N_w\Delta_t \approx qT_o$, where $q$ is an integer, then the RMSF is found to preserve the function, see Appendix~\ref{App:Sine-RMSF}.

In the next section, using a simulated data stream, we demonstrate the application of RMSF and its advantage over RASF.

\section{Efficacy of running median filter in removing baselines }\label{Sec:Sim}
In the introduction, Sec.~\ref{Sec:Intro}, we compared the efficacy of RMSF with that of RASF in removing a slowly varying baseline, Fig.~\ref{Fig:Filter-Plots}. In this section, we present technical details which went into that simulation and present additional insights. 

The simulated time series in Fig.~\ref{Fig:Filter-Plots}, denoted here by $\{x_j\}$, is a sum of a normally distributed white noise stream~$\{n_j\}$, a Gaussian signal~$\{g_j\}$, and a slowly varying sinusoidal baseline variation~$\{b_j\}$, i.e., $x_j=n_j+g_j+b_j$. The time grid, $t_j=j\Delta_t$, is in units of seconds, where the index $j=\overline{1,N}$, the time step $\Delta_t=1$~s and sample size $N=1000$. We will use this uniform sampling grid throughout {this article}. The white noise stream $n_j\sim \mathcal{N}\left(\mu,\sigma^2\right)$ has a mean $\mu=0.5$ and standard deviation $\sigma=0.1$. The Gaussian signal,
\begin{align}
    g_j=2\exp\left[{-\frac{\left(t_j-t_s\right)^2}{2\tau^2}}\right]\,,\label{Eq:Gaussian}
\end{align}
is centered at $t_s=500$~s, with the signal duration $\tau=20$~s. Here and below we pick $t_s$ to coincide with a grid point. The baseline variation is given by the function $b_j=0.5\sin\left[2\pi t_j/T_o\right]$ with the period $T_o=2\pi N\Delta_t/4\approx1600$~s, which is much longer than the signal duration $\tau$. 

In Fig.~\ref{Fig:Filter-Plots}(a), we compare the RMSF processed data with the original data and the Gaussian signal~\eqref{Eq:Gaussian}. We observe that the RMSF removes the slowly varying baseline without significantly distorting the signal. The results of the RMSF application depends on the choice of the running window size. There is a trade off: a wider running window truncates the processed data more, whereas, a narrower running window distorts the transient signal more, see Sec.~\ref{Sec:Sig-RMSF}. Hence, one needs to choose an optimum size of a running window so as to preserve the signal to the maximum extent without loosing too many data points. In Fig.~\ref{Fig:Filter-Plots}, we used the running window of $N_w=161$ points that was found to satisfy the above criteria through numerical experimentation. We will return to the question of choosing optimal window size in Sec.~\ref{Sec:Sig-RMSF}.

Now we discuss the efficacy of RMSF over RASF (median over average subtraction). The result of applying the RASF, Eq.~\eqref{Eq:mean-sub}, to the data stream, is shown in Fig.~\ref{Fig:Filter-Plots}(b). Here we keep the same window size of $N_w=161$ points as in the RMSF simulation. In the RASF processed data, we observe a noticeable distortion of the signal, even though the baseline variation is removed. There are two dips, one on each side of the signal in the RASF-processed data along with a vertical displacement of the signal. Such dips in the RASF-processed data are proportional to the signal amplitude, whereas in the RMSF-processed data the dips are negligible and independent of the signal strength for signals above the noise level~\cite[Chap. 5, Sec. 5.4.4]{Gregory}. 

Indeed, following~\cite[Chap. 5, Sec. 5.4.4]{Gregory}, consider stationary noise data $\{n_j\}$ in a window. Let the median and the average of the noise over the window be $n_{\mathrm{med}}$ and $n_{\mathrm{avg}}$, respectively. 
Suppose a spike of strength $S>0$ is added at a particular time,~$t_j$, within the window. Such a spike is a limiting case of a Gaussian signal~\eqref{Eq:Gaussian} when the signal duration $\tau$ is much smaller than the time step $\Delta_t$ with $t_s=t_j$. In this limit, the signal appears as a spike. We require that $S$ is comparable to, or larger than the noise standard deviation. If the noise $n_j$ at $t_j$ is greater than $n_{\mathrm{med}}$, then after adding the signal $S$, the median remains unchanged, see~\eqref{Eq:Median}. However, if $n_j<n_{\mathrm{med}}$, then when $n_j+S>n_{\mathrm{med}}$, the data value moves to the other side of $n_{\mathrm{med}}$ in the sorted data stream, increasing the median. Upon further increase in the signal strength $S$, the median remains unchanged, see Sec.~\ref{Sec:Def}. Therefore, the dips in the case of RMSF are independent of the signal strength when the signal duration is much smaller than the window size. In contrast, the {\it average} value always increases by $S/N_w$ upon adding signal to a noise sample, see Eq.~\eqref{Eq:Mean}.
Hence, the dips and distortion of the signal in the RASF processed data depend on the signal strength.

Thus, based on the analysis and the discussions done in this section, RMSF is more effective than RASF in removing slowly varying backgrounds in a data stream. As a side note, in the field of image processing it is well known that median-based spatial filters (in 2-dimensions) preserve edges, while being able to despeckle images (i.e., remove or ignore outliers)~\cite{Hammond}. On the other hand, mean-based filters blur edges and distort images.  This is analogous to the one-dimensional problem discussed here.

\section{Fidelity of RMSF with a Gaussian transient}\label{Sec:Sig-RMSF}

To be effective, data processing filters must possess the important property of not substantially distorting signals of interest in a data stream. 
Any transient is a non-monotonic function by definition, and the RMSF  does not fully eliminate it (see discussion in Sec.~\ref{Sec:Def}); if a given window contains the {extremum of the transient}, the filtered data is non-zero. We are interested in the regime when the signal of interest is preserved.  We will show that this happens when the running window size $N_w$ is much larger than the transient duration $\tau$.  For concreteness, in this section, we explore the effect of the RMSF on a generic Gaussian transient~\eqref{Eq:Gaussian}.

\begin{figure}[h!]
    \centering  \includegraphics[width=1.0\columnwidth]{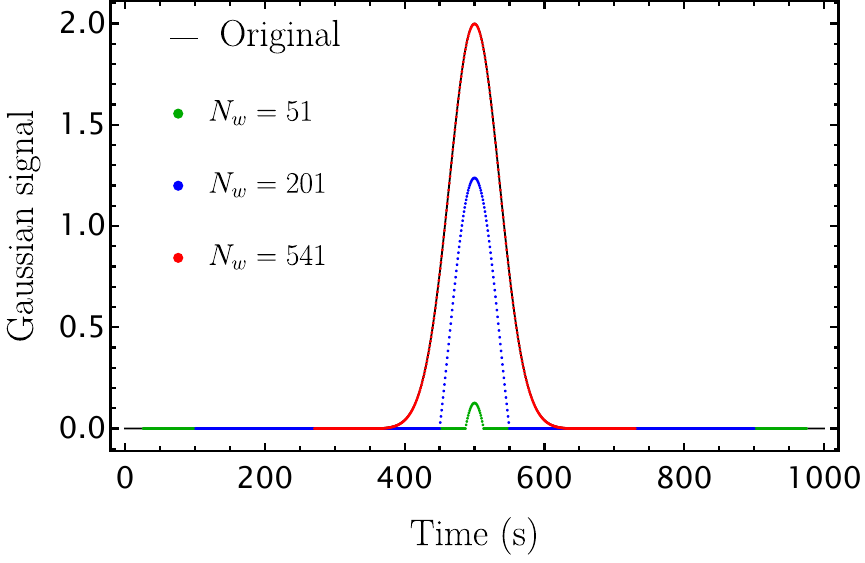}
    \caption{Transformation of a Gaussian signal~\eqref{Eq:Gaussian}, upon application of the running median subtraction filter (RMSF) with different running window sizes $N_w$. The signal fidelity improves with larger window size.}\label{Fig:Sig-Plot}
    \end{figure}

To demonstrate this, we apply RMSF to a Gaussian signal~\eqref{Eq:Gaussian} centered at $t_s=500$~s, with signal duration $\tau=36$~s and a peak value of 2. As in Sec.~\ref{Sec:Sim}, we keep the same uniform time grid of $N=1000$ points with a step $\Delta_t=1$~s. Fig.~\ref{Fig:Sig-Plot} shows the effect of varying window size $N_w$ on the filtered signal. We see that the signal is significantly distorted when $N_w\Delta_t$ is small compared to the signal width. In other words, the RMSF yields better signal fidelity with larger window size. 

To quantify the signal fidelity, we define a root mean square (rms) deviation of the RMSF processed signal $\Tilde{g}_j$, from the original signal $g_j$,
\begin{align}
\begin{split}
   &(\Delta g_{\text{rms}})^2 \\ & =\frac{1}{5N_{\tau}+1}\sum_{j=j_s-2.5N_{\tau}}^{j_s+2.5N_{\tau}}\left(g_j-\Tilde{g}_j\right)^2,  
\end{split}
  \label{Eq:RMS}
\end{align}
where $N_{\tau}=\tau/\Delta_t$ is the number of points spanned by the signal duration $\tau$ and $j_s=t_s/\Delta_t$. This rms deviation is computed over the bulk of the transient signal. 
To quantify ``bulk", we define the signal width $t_w$ as the time span between two points in a Gaussian function at which the signal drops by 95\% from its peak value,
\begin{align}
t_w=2\tau\sqrt{2 \ln \left(\frac{1}{0.05}\right)}\approx 5\tau\,.
\end{align}
$t_w$ corresponds to $5N_{\tau}$ points used in the definition of rms deviation~\eqref{Eq:RMS}. We further define a fractional error as the ratio of the rms deviation to the signal peak value $g_s\equiv g_{j_s}$, 
\begin{align}
    \varepsilon=\frac{ \Delta g_{\text{rms}}}{|g_s|}\,.\label{Eq:Rel-Error}
\end{align}

\begin{figure}[h!]
    \centering  \includegraphics[width=1.0\columnwidth]{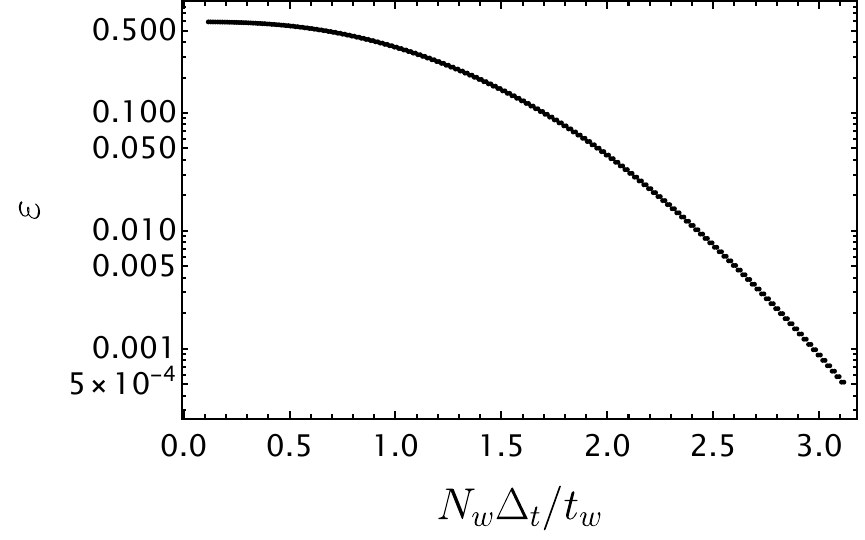}
    \caption{Fractional error $\varepsilon$ in the RMSF filtered Gaussian signal decreases with the increasing relative window size $N_w\Delta_t/t_w$.}\label{Fig:RMS}
    \end{figure}

Fig.~\ref{Fig:RMS} shows our computed  fractional error $\varepsilon$ as a function of window size normalized by the signal width, $N_w\Delta_t/t_w$, termed here as relative window size. This figure quantifies the RMSF signal fidelity as a function of the relative window size for a Gaussian signal. The fractional error decreases with increasing relative window size and it drops below 0.1\% for $N_w\Delta_t/t_w\geq3$. 

To explain this behavior, we focus on how the peak value $g_s$ of the Gaussian is transformed by RMSF. The value $\tilde{g}_s$ comes from the windows centered on $t_{j=s}$. We consider a sequence of such windows of increasing size $N_w$ and show that the error $\delta_g=|g_s-\tilde{g}_s|$ becomes smaller, c.f. Fig.~\ref{Fig:Sig-Plot}.  If $N_w=1$, then the median is $g_s$, so that $\tilde{g}_s=0$ resulting in a large error, $\delta_g=g_s$. Further, $N_w=3$ window contains data $\{g_{s-1},g_s,g_{s+1}\}$, where $g_{s-1}=g_{s+1}$, due to the symmetry of the Gaussian, and $g_{s-1}<g_s$. Specifically, from~\eqref{Eq:Gaussian}, $g_s=2$ and $g_{s\pm 1} \approx1.9975$, with a window median of $1.9975$. Thus the RMSF processed value $\tilde{g}_s$ becomes $2-1.9975=0.0025$. Likewise, the data in a window of size $N_w=5$ are $\{g_{s-2},g_{s-1},g_s,g_{s+1},g_{s+2}\}$, where $g_{s-2}<g_{s-1}<g_s$ and $g_{s-2}=g_{s+2}$ due to the symmetry of Gaussian. Now using~\eqref{Eq:Gaussian}, we have $g_{s\pm2}\approx1.9900$, with the median of the $N_w=5$ array being $1.9975$. Similarly, the medians for $N_w=7,9,11~\text{and}~13$ windows are $1.9776,1.9776,1.9604~\text{and}~1.9604$, respectively. We see that the median moves further away into the tails of the Gaussian with growing window size. This reduces the difference between the original data and the RMSF processed data, $\delta_g$, according to~\eqref{Eq:med-sub}, thus decreasing the fractional error, resulting in higher fidelity. We also observe that we get the same median value for two successive windows, e.g., $N_w=7$ and $N_w=9$. Such pairwise equal medians lead to repeating doublet degeneracy in fractional errors apparent in Fig.~\ref{Fig:RMS}. 

We expect a similar improvement in fidelity with increasing relative window size for any transient signal. We must require that the running window size is much larger than the transient signal duration. In the following section we extend this analysis to a transient signal characterizing exotic physics modality in multi-messenger astronomy.

\section{Fidelity of  RMSF with an exotic low-mass field transient}\label{Sec:Sig-ELF}
In the previous section we dealt with a Gaussian signal; in this section we investigate the effect of RMSF on a signal of interest generated by hypothetical exotic low-mass fields (ELFs). ELFs are plausibly emitted by intense astrophysical events such as binary black hole and neutron star mergers \cite{Dailey2020,khamis2024}, extending multi-messenger astronomy to exotic physics modality. Multi-messenger astronomy is the coordinated observation of different classes of signals that originate from the same astrophysical event. 
ELF signals could be classified as oscillating transients. In our GPS.ELF collaboration~\cite{Sen_2024}, such signals are being sought in the data streams from atomic clocks aboard the satellites of Global Positioning System (GPS). A typical data stream contains slowly varying backgrounds which need to be removed while preserving the ELF signal of interest. This serves as a motivation for this section.

An ELF signal is a product of a Gaussian envelope and an oscillatory factor. While the ELF signal is chirped, here we model it as
\begin{align}
\begin{split}
    h_j & = A \exp{\left[-\frac{{\left(t_j-t_s\right)^2}}{2\tau^2}
\right]} \\ & \times\cos\left[\frac{2\pi}{T}\left(t_j-t_s\right)\right]\,,   
\end{split}
 \label{Eq:ELF}
\end{align}
where $T$ is a fixed oscillation period. The peak amplitude~$A$ and the duration $\tau$ of the signal depends on the physics of the ELF emission and propagation~\cite{Dailey2020}; we take $A=2$, the same as in our model Gaussian signal~\eqref{Eq:Gaussian}. We also retain the same time grid, signal duration $\tau$, and $t_s$ as in Sec.~\ref{Sec:Sig-RMSF}.

\begin{figure}[h!]
    \centering  \includegraphics[width=1.02\columnwidth]{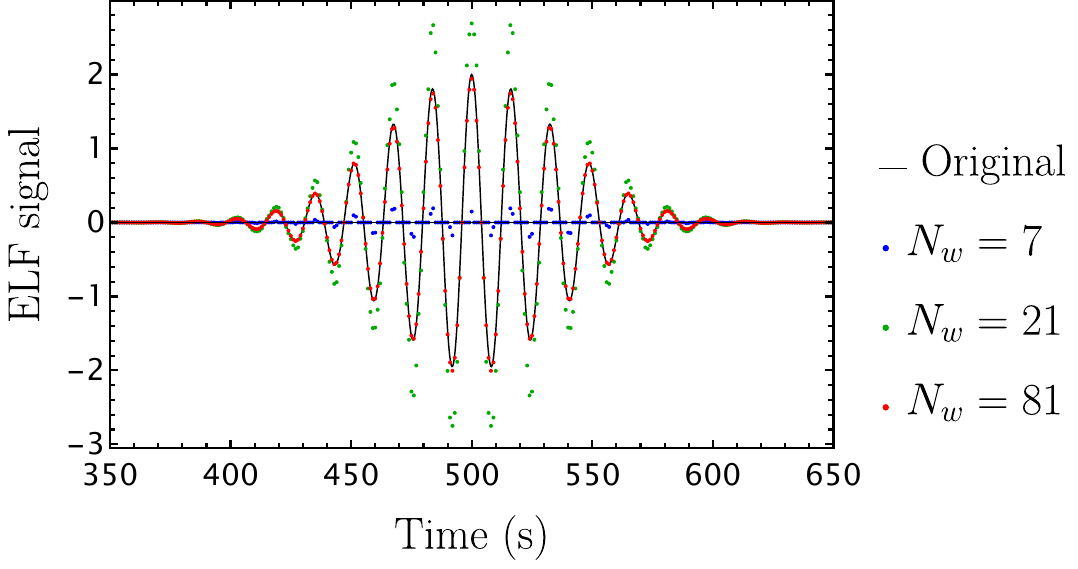}
    \caption{{Transformation of the ELF signal~\eqref{Eq:ELF} with a period $T=16.3$~s, upon application of the running median subtraction filter (RMSF) with different running window sizes $N_w$. The signal fidelity improves when $N_w\approx qT$, where $q$ is an integer. As we reduce the window size, the signal becomes suppressed except near the points of extrema. The RMSF-processed signal increases when $1.5T \lesssim N_w\Delta_t <2T$, due to the sinusoidal part in Eq.~\eqref{Eq:ELF}, per Sec.~\ref{Sec:Def}. Here, we kept the same time grid, signal duration $\tau$ and signal peak time $t_s$ as those in Fig.~\ref{Fig:Sig-Plot}.}}\label{Fig:ELFSig-Plot}
    \end{figure}

The ELF signal~\eqref{Eq:ELF} is governed by two characteristic time scales: the signal duration $\tau$ and the oscillation period $T$. The RMSF fidelity is expected to depend on these two time scales along with the running window duration $N_w\Delta_t$. In Fig.~\ref{Fig:ELFSig-Plot}, taking $T\ll\tau$, we compute a transformed signal $\Tilde{h}_k$ for three different window sizes $N_w$. The ELF signal~\eqref{Eq:ELF} has a Gaussian envelope and a sinusoidal part. As for a Gaussian signal, we find that the signal fidelity improves as window size increases. However, owing to the presence of a sinusoidal part, the ELF signal~\eqref{Eq:ELF} is maximally preserved when the RMSF window duration $N_w\Delta_t$ is approximately integer multiples of $T$, see Sec.\ref{Sec:Def}. For windows narrower than the period $T$, the signal tends to get suppressed except near the points of extrema, per Sec.~\ref{Sec:Def}. Also, when $1.5 \lesssim N_w\Delta_t <2T$, we see an amplification of the signal. It is due to the sinusoidal part of the ELF, see Sec.~\ref{Sec:Def}. 

\begin{figure}[h!]
    \centering  \includegraphics[width=1.0\columnwidth]{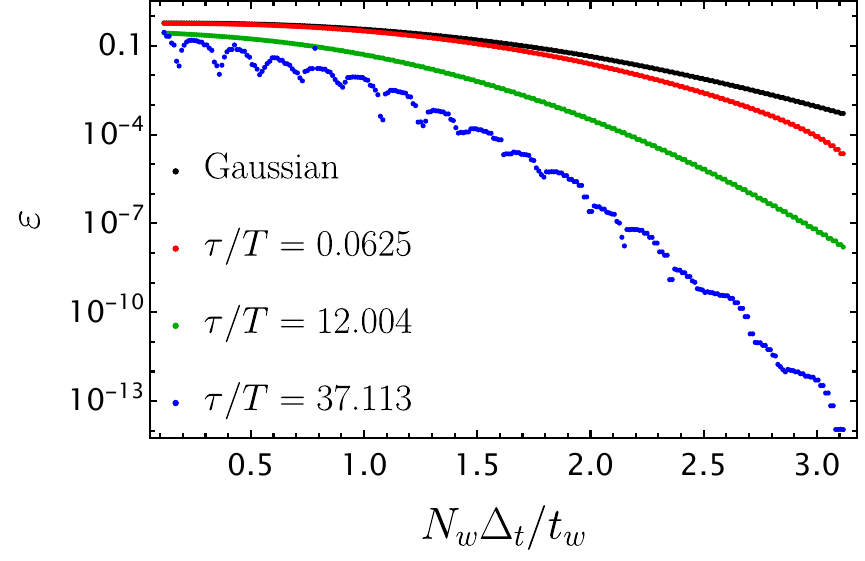}
    \caption{Fractional error $\varepsilon$ in the running median subtraction filter processed exotic low-mass field (ELF) signal~\eqref{Eq:ELF} as a function of the relative window size. The ELF signals~\eqref{Eq:ELF} have different oscillation periods, $T=576~\mathrm{s},~2.999~\mathrm{s}$ and $0.970$~s and a fixed duration $\tau=36$~s, $t_s=500$~s.}\label{Fig:Log-err}
    \end{figure}

{To quantify the RMSF fidelity for the ELF signal}, we employ Eqs.~\eqref{Eq:RMS} and \eqref{Eq:Rel-Error}, but with $g_j\rightarrow h_j$. Fig.~\ref{Fig:Log-err} shows the fractional error $\varepsilon$ as a function of $N_w\Delta_t/t_w$ for different oscillation periods $T$ of the ELF signal. We observe that as $T$ increases for a fixed $\tau$, the fractional error curve for the ELF signal approaches the error curve for the Gaussian signal $g_j$ (black curve). This is due to the fact that, as $T\rightarrow \infty$, the ELF signal recovers non-oscillating Gaussian signal, $h_j\rightarrow g_j$. The error for the ELF signal with finite $T$ is always smaller than that for the Gaussian error.

As the ELF signal period $T$ decreases, so does the fractional error $\varepsilon$. This is because for a given window size $N_w$, a shorter period implies that the ELF signal width becomes narrower. Indeed the ELF signal width, $t^{\mathrm{ELF}}_w$, is approximately the time interval between the first zeroes of $h_j$ straddling the peak, which gives $t^{\mathrm{ELF}}_w=T/2$. Then a smaller period implies a smaller signal width compared to the window size, resulting in a larger relative window size $N_w\Delta_t/t^{\mathrm{ELF}}_w$. Since with larger relative window size the error drops, we observe that an ELF signal of a lower period $T$ is perturbed less by the filter. 

Another feature of the plot in Fig.~\ref{Fig:Log-err} is that there are certain periodicities in the fractional error curve of the ELF signal. This can be traced to the oscillatory factor in the signal. A larger period gives rise to a larger curvature of the relative error curve, whereas, a smaller $T$ leads to a periodic repetition of smaller chunks or the discontinuous jumps in the relative error curves of Fig.~\ref{Fig:Log-err}.

A generic data stream contains a normally distributed white noise, a signal component and a baseline variation. So far we have studied elimination of background function and how preserve the signal of interest. In the following section, we explore the RMSF effect on a normally distributed white noise, the remaining component of a generic data stream. 


\section{Application of RMSF to a normally distributed white noise stream}\label{Sec:Norm-RMSF}

What happens to a white noise stream when the running median subtraction is applied? To answer this question, we consider normally distributed white noise streams and apply RMSF.  We will demonstrate that in the limit of large windows, $N_w \gg 1$, the running median filter does not affect the statistical properties of the filtered noise, except for removal of the data mean. The removal of the mean is similar to the RMSF elimination of constant functions, see Sec.~\ref{Sec:Def}.

What do we mean by statistical properties of the original and filtered data being the same?
Suppose we  generate realizations or runs of random data drawn from some distribution. Given the stochastic nature, these realizations are not identical, however their distribution functions are. In practice, to show that the their underlying distribution functions are the same,  we need to estimate different statistics such as the data mean, variance, and kurtosis. If these statistics agree within the sampling error, then the underlying distribution functions are identical. The same argument applies to auto-correlation functions. We will demonstrate that the filtered white noise is not colored by the filter.

We carry out numerical experiments by generating normally distributed white noise runs $\{n_j\}$ of $N=500$ points, mean $\mu=5$ and standard deviation $\sigma =10$. We estimate the sample variance $\hat{\sigma}^2$, 
skewness $\hat{\mathrm{s}}$
and kurtosis $\hat{\kappa}$ as 
\begin{align}
    &\hat{\sigma}^2=\frac{1}{N-1}\sum_{j=1}^N\left(n_j-\hat{\mu}_1\right)^2\,,\\
   &\hat{\mathrm{s}}=\frac{\hat{\mu}_3}{\hat{\sigma}^{3}}\,,\\
    &\hat{\kappa}=\frac{\hat{\mu}_4}{\hat{\sigma}^4}\,,
\end{align}
where for $r \ge 2$ we define $\hat{\mu}_r$ to be the $r^{\text{th}}$ centered sample moment, 
\begin{align}
\hat{\mu}_r=\frac{1}{N}\sum_{j=1}^N\left(n_j-\hat{\mu}_1\right)^r.
\end{align}
Here $\hat{\mu}_1$ is the sample mean of the stream~$\{n_j\}$ (we skip index $1$ and refer to $\hat{\mu}_1$ as simply $\hat{\mu}$ for consistency with the literature) and $\hat{\mu}_2$, $\hat{\mu}_3$
and $\hat{\mu}_4$ are the second, third and fourth sample moments respectively. Note that $\hat{\sigma}^2$ is almost the same as $\hat{\mu}_2$, only we divide by $N-1$ instead of $N$. 

 In our comparisons of the enumerated statistics, we also need to quantify  sampling errors in their determination. 
 For example, according to the {\it central limit theorem}, a data stream of IID random variables with a sample size $N$ and a finite standard deviation $\sigma$ and sample mean $\hat{\mu}$ asymptotically follows a normal distribution with a standard error of $\sigma/\sqrt{N}$~\cite[Chap. 5, Sec. 5.9]{Gregory}. If the original data stream was normal, then the sample mean would have precisely a normal distribution as well. Moreover, when the data stream consists of IID normally distributed random variables, then $(N-1)\hat{\sigma}^2/\sigma^2$ is distributed as a chi-squared distribution with $N-1$ degrees of freedom~\cite[Theorem 5.3.1(c)]{Casella}. As shown in~\cite[Example 7.3.3]{Casella} that, under these conditions, the sample variance $\hat{\sigma}^2$ has the standard error
 \begin{equation}
 \label{eq:sample-var-std}
\hat{\sigma}^2\sqrt{\frac{2}{N-1}}.  
 \end{equation}
Similarly, the sample 
skewness and sample kurtosis are distributed asymptotically normally  with a sample error of 
$\sqrt{6/N}$ and
$\sqrt{24/N}$, respectively~\cite[Chap. 10, Exercises 10.26 and 10.27]{kendall}.

In our numerical experiment, the average value of the sample means over 100 runs is $\approx 5.01$, in agreement with $\mu=5$ of the underlying distribution.  The average of the sample mean after applying RMSF is $\approx -0.015$, with the sampling error of $\sigma/\sqrt{100\, N}\approx 0.045$, indicating that the mean of the filtered data is consistent with zero.  As our numerical experiment demonstrates, the median subtraction filter removes the mean, i.e., it removes the constant background as expected. Indeed, consider applying RMSF to a normally distributed continuous variable $x$. The expectation value of the RMSF-processed variable is $\langle \tilde{x} \rangle= \langle x - \mathrm{median} \rangle  =\langle x \rangle -\mathrm{median}$. For a symmetric distribution, $\langle x \rangle = \mathrm{median}$, see Sec.~\ref{Sec:Def}. Hence, $\langle \tilde{x} \rangle =0$. 

Next, we calculate the average sample variance over the 100 runs as $\hat{{\sigma}}_{\mathrm{av}}^2=\sum_{i=1}^{100}\hat{\sigma}^2_i/100$. From our simulation, we find  the difference in the average sample variance of the original noise stream and the RMSF processed noise, 
$$
\delta_{\sigma^2}=\bigl|\hat{{\sigma}}_{\mathrm{av}}^2-\hat{\tilde{\sigma}}_{\mathrm{av}}^2\bigr|\approx 0.54.
$$ 
Using Eq.~\eqref{eq:sample-var-std}, it is easy to show that the standard error of sample variance is given by 
$\hat{\sigma}^2\sqrt{2/(100(N-1))}\approx 0.63$. Thereby, the difference $\delta_{\sigma^2}$ is within the sampling error of the variance and our simulation implies that the filter preserves the variance of the original data, in the limit of large window size. See Appendix~\ref{App:Var-RMSF} for a proof of this finding.

As a numerical test of higher-moment statistics, we also determined the 
average sample skewness over 100 runs, $\hat{{\mathrm{s}}}_{\mathrm{av}}=\sum_{i=1}^{100}\hat{\mathrm{s}}_i/100$. The difference in the average skewness of the original and the filtered noise is
\begin{align*}
\delta_{\mathrm{s}}=\bigl|\hat{{\mathrm{s}}}_{\mathrm{av}}-\hat{\tilde{\mathrm{s}}}_{\mathrm{av}}\bigr|\approx  0.002.
\end{align*}
The corresponding sampling error for skewness is $\sqrt{6/100N}$, which is $\approx 0.01$ for this simulation. Hence, the difference $\delta_{\mathrm{s}}$ is within the sampling error of skewness, which implies that RMSF preserves the skewness of the original data. Now, the 
average sample kurtosis over 100 runs is $\hat{{\kappa}}_{\mathrm{av}}=\sum_{i=1}^{100}\hat{\kappa}_i/100$. We computed the difference in the average kurtosis of the original and the filtered noise 
\begin{align*}
\delta_{\kappa}=\bigl|\hat{{\kappa}}_{\mathrm{av}}-\hat{\tilde{\kappa}}_{\mathrm{av}}\bigr|\approx 0.06.
\end{align*}
The corresponding sampling error is $\sqrt{24/100N}$, which is $\approx 0.02$ for this simulation. The distribution of the kurtosis is more fat tailed than that of the sample mean, variance and skewness, as the distribution of $\hat{\mu}_4$ has fatter tails than that of $\hat{\mu}$, $\hat{\mu}_2$ and $\hat{\mu}_3$. Hence, for $\delta_{\kappa}$ to be comparable to the sampling error, it is expected to require a larger number of runs and sample size. 

\begin{figure}[h!]
    \centering  \includegraphics[width=1\columnwidth]{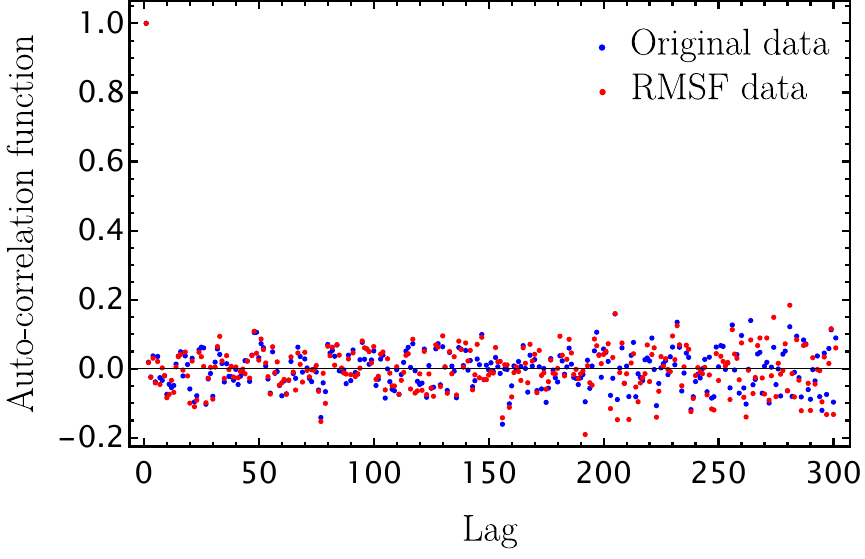}
    \caption{Comparison of the ACF $\hat{\rho}(l)$ of the original white noise and of the RMSF-transformed noise. It can be seen that the filtered white noise remains uncorrelated with $\hat{\rho}(l)$  being comparable to that of the original stream. Here the window size $N_w=51$ and the sample size $N=500$.}\label{Fig:ACF}
    \end{figure}

We now check if RMSF induces any correlation within the filtered initially uncorrelated noise stream. Does the filter color the original white noise? To this end we calculate the auto-correlation function (ACF). The ACF of the stream, ${\rho}(l)=\mathrm{Corr}({n}_j, {n}_{j+l})$, quantifies the correlation between $n_j$ and $n_{j+l}$ which depends on the lag $l$.
White noise ACF is ${\rho}(l)=\sigma^2 \delta_{l,0}$, where $\sigma^2$ is the noise variance.
The estimated ACF is~\cite[Chap. 5, Sec. 5.13.2]{Gregory} 
\begin{align}
    \hat{\rho}(l)=\frac{\sum_{j=1}^{N-l}\left(n_j-\hat{\mu}\right)\left(n_{j+l}-\hat{\mu}\right)}{\sum_{j=1}^{N}\left(n_j-\hat{\mu}\right)^2}.
    \label{eq:estimate-ACF}
\end{align}
This statistic~\eqref{eq:estimate-ACF} is a consistent estimate of the true ACF: $\hat{\rho}(l) \to \rho(l)$ {\it almost surely} (with probability 1) as $N \to \infty$ for each $l = 1, 2, \ldots$
In Fig.~\ref{Fig:ACF}, we compare ACFs of the white noise stream before and after the RMSF application. Clearly, our simulation implies that the RMSF does not induce any noticeable correlation. See Appendix~\ref{App:ACF-RMSF} for the proof of this finding. This explains the ACF comparison of the original and filtered noise in Fig.~\ref{Fig:ACF}.

To summarize, the RMSF removes the mean of a normally distributed white noise while preserving its variance and higher order moments in the limit of large windows. In addition, RMSF does not color the white noise stream (does not induce any significant correlation in the filtered data).

\section{Conclusions}
\label{Sec:Conclusions}
The running median subtraction filter (RMSF) is a robust tool for removing slowly varying baselines. In this tutorial, through numerical experimentation and mathematical proofs, we demonstrated several important properties of this filter. For sufficiently wide running windows, the RMSF does not perturb transient signals of interest, such as the oscillating transients characteristic of exotic physics modalities in multi-messenger astronomy. When applied to white noise, the RMSF removes the mean while preserving the variance and higher-order moments of the original noise. Additionally, the filter  does not induce any significant correlation in the filtered data.

Practitioners must pay attention to choosing the running window size: the window must be wider than the width of the transient signal of interest, while its size must be smaller than the characteristic time scale of undesired background clutter.
As a consequence, one limitation of the RMSF is that the practitioner must assume something about the time duration of the signal of interest, when there may be little theoretical guidance. Overall, we find the RMSF to be an efficient choice for analyzing data with slowly varying baselines. It is largely insensitive to outliers and can preserve both transient signals and the stochastic characteristics of data streams.

\begin{acknowledgments}

This work was supported in part by the U.S. National Science Foundation grant PHY-2207546, Heising-Simons Foundation grant~2022-3916, and NASA grant~NV-80NSSC23M0104.

\end{acknowledgments}

\bigskip

{\bf \center AUTHOR CONTRIBUTIONS\endcenter}

\medskip
All provided methodology, review, and editing.    A.P.S. wrote the initial draft, wrote software, and conducted numerical experiments. A.S. performed formal statistical analysis. G.B. conceptualized the application of robust statistics and provided supervision. A.D. conceptualized the work and provided project administration.

\bigskip

{\bf \center AUTHOR DECLARATIONS\endcenter}

\medskip
The authors have no conflicts to disclose.

\appendix 
\section{Application of RMSF to a sinusoidal function}\label{App:Sine-RMSF}

In this section, we consider the median of a continuous sinusoidal function, rather than the median of a discrete data following this sine function. We do this for the sake of simplicity, and to make exposition more elegant. The median of a sine function $f : [a, b] \to \mathbb R$ defined as $f(t) = A\sin\left(2\pi t/T_o\right)$ (with period $T_0$) can be defined as the median of the random variable $X := f(U)$, where $U$ is a uniform random variable on $[a, b]$. If $t_0 = a < t_1 < t_2 < \ldots < t_n = b$ is equally distributed data, so that $t_k - t_{k-1} = (b-a)/n$, then the median of the data $f(t_0), \ldots, f(t_n)$ converges to the median $m$ of this function $f$. This follows from the fact that the random variable $X$ takes value $m$ with probability zero. See more on this topic in~\cite{Bivens}.

Let a random variable be $X = \sin \theta$, where $\theta \sim \mathrm{Uniform}\left[ -\pi/2,\pi/2\right]$. The mean and median of the variable $\theta$ are zero. The sine function increases from $-1$ to 1 in the range $\left[-\pi/2,\pi/2\right]$. The probability of the variable $X$ to be greater or less than the median is $1/2$, as the median is the value which splits the area under the PDF in halves. Thus, the median of $X$ is the sine of median of $\theta$, which is zero.

Similarly, if $\theta$ is restricted to a sub-interval $\left[a,b\right] \leq \left[-\pi/2,\pi/2\right]$, then its median is $(a+b)/2$ and the median of the variable $X=\sin\theta$ is $\sin\left(\frac{a+b}{2}\right)$.

Now applying the RMSF to a continuous sine function $f(t)=A\sin\left(2\pi t/T_o \right)$ within a window $\left[-D,D\right]$, we get
\begin{align}
\begin{split}
    \tilde{f}\left({t'}\right) &= f\left( {t'}\right) - m\left( {t'}\right);\\
    m\left( {t'}\right) &=\mathrm{Median}\left[ \{ f(t)\}_{|t-{t'}|\leq D}\right].\, 
\end{split}
\end{align}
Now, depending on the sign of $m(t')$ the filtered result will be amplified, dampened or preserved. We have $m(t')>0$ if and only if the Lebesgue measure of $\{|t-t'|\leq D~|~\sin \left( 2\pi t/T_o \right)>0 \}>D$, which is half the length of $\left[ t'-D,t'+D \right]$. After solving the inequality $ \sin \left( 2\pi t/T_o \right)>0$, we get that this is true if and only if one of the two cases hold:
$$
\begin{cases}
\sin\left( 2\pi t'/T_o \right) > 0,\quad t' \in \mathbb D_+;\\
\sin\left( 2\pi t'/T_o \right) < 0,\quad t' \in \mathbb D_-.
\end{cases}
$$
We define new notation
\begin{align*}
\mathbb D_+ &:= (0, T_o/2)\cup (T_o, 3T_o/2)\cup 2T_o, 5T_o/2)\cup\ldots\\
\mathbb D_- &= (T_o/2, T_o)\cup(3T_o/2, 2T_o)\cup(5T_o/2, 3T_o)\cup\ldots\\
\mathbb D_0 &= \{qT_o/2\mid q = 1, 2, \ldots\}
\end{align*}
Conversely, $m(t') < 0$ if and only if one of two cases hold: 
$$
\begin{cases}
\sin\left( 2\pi t'/T_o \right) < 0,\quad t' \in \mathbb D_+;\\
\sin\left( 2\pi t'/T_o \right) > 0,\quad t' \in \mathbb D_-.
\end{cases}
$$
Finally, $m(t') = 0$ if and only if the Lebesgue measure of $\{|t - t'| \le D\mid \sin\left( 2\pi t/T_o \right) > 0\}$ is equal to $D$, which happens when $D \in \mathbb D_0$. Thus, $f(t')$ and $m(t')$ have the same sign for all $t'$ if $D \in \mathbb D_+$; $f(t')$ and $m(t')$ have the same sign for all $t_k$ if $t' \in \mathbb D_-$; and $m(t') = 0$ for all $t'$ if $t' \in \mathbb D_0$. In other words, the filter dampens the signal for $D \in \mathbb D_+$, and amplifies it if $D \in \mathbb D_-$. The filter does not change the signal if $D \in \mathbb D_0$. 

But the amplification or dampening is not by the same pre-factor $C$. We propose a theorem: There exists a value $C>0$ such that the RMSF filtered result $\tilde{f}(t)$, for large enough window width $D$, applied to the sinusoidal function $f(t)=A\sin\left(2\pi t/T_o \right)$, satisfies the relation $|\tilde{f}(t)| \le C/D~\forall~t\in \mathbb R$. 
\paragraph{Proof:} Let $K=\left[D/T_o\right]$, the integer part of $D/T_o$. Hence, there are $K$ full periods, denoted by $P_1,P_2,...,P_K$ inside the window $\left[t-D/2,t+D/2\right]$ and part $P_0$ of another period. For the median $m$ of $f(t)$ in $\left[t-D/2,t+D/2\right]$, the Lebesgue measure (sum of interval lengths) for $\{u \in P_j\mid f(t) \le m\}$ over these $K$ periods is 
$$
Q = \frac{T_o}2 + \frac{T_o}{\pi}\sin^{-1}\left(\frac{m}{A}\right).
$$
Assume $L$ is the Lebesgue measure of $\{u \in P_0\mid f(t) \le m\}$. Of course, $0 \le L \le T_o$. Since $m$ is the median of $f$ upon $[t-D/2, t+D/2]$, the overall set $\{u \in [t - D/2, t+D/2] \mid f(t) \le m\}$ has Lebesgue measure $D/2$. But this is the sum of these $K+1$ terms, which is $KQ + L$. Thus $KQ + L = m/2$. Combining this and solving for $m$, we get:
$$
m = A\sin\left[\frac{\pi}{T_o}\left(\frac{D}{2K} - \frac{L}{K} - \frac{T_o}2\right)\right].
$$
Since $K = [D/T_o]$, we have: $0 \le D/K - T_o \le T_o$. Also, $0 \le L \le T_o$. Thus 
$$
\left|\frac{D}{2K} - \frac{L}{K} - \frac{T_o}2 \right| \le \frac{T_o}{2K}.
$$
Since $|\sin(x)| \le |x|$ for all real $x$, we get: 
$$
|m| \le A(\pi/T_o) \cdot (T_o/2K).
$$
Remembering that $K = [D/T_o]$, we complete the proof.

\section{Preservation of white noise sample variance under RMSF}\label{App:Var-RMSF}
Consider the series $\{n_j\}$, with $j=\overline{1,N}$, which is normally distributed with a mean $\mu$ and variance $\sigma^2$. The RMSF-processed result~\eqref{Eq:med-sub}, is
\begin{align}
\Tilde{n}_k=n_k-\mathrm{Median}\left[\{n_j\}_{j=k-M}^{k+M}\right]\,, \label{eq:RMSF-k}
\end{align}
where $M=(N_w-1)/2$ and $N_w$ is the window size. 
The variance of filtered series $\{\tilde{n}_k\}$ is
\begin{align}
\begin{split}
    \tilde{\sigma}^2 & = \sigma^2 + \sigma^2_{\mathrm{median}} \\ & -2 \,\mathrm{Cov}\left( n_k,\mathrm{Median}\left[\{n_j\}_{j=k-M}^{k+M}\right] \right)\,, \label{Eq:Var-Med-0}
\end{split}
   \end{align}
where $\sigma^2_{\mathrm{median}}$ is the variance of the median. For sufficiently large windows, the sample median is approximately normally distributed with a variance~\cite[Chap. 10, Sec. 10.10]{kendall}~\cite[Chap. XIII, Sec. 13.13]{kenney1951}
\begin{align}
    \sigma^2_{\mathrm{median}} \approx \frac{\pi\sigma^2}{2N_w}\,.\label{Eq:Var-Med}
\end{align}
A remarkable result was proved in~\cite{Rinott1994} (for normal data stream) on the covariance between $n_k$ and the median:
\begin{align}
\label{Eq:Cov-nk-med}  
\begin{split}
&\mathrm{Cov}\bigl( n_k,\mathrm{Median}\bigl[\{x_j\}_{j=k-M}^{k+M}\bigr] \bigr) \\ & = \sum\limits_{j=k-M}^{k+M}
\mathrm{Cov}\bigl(n_k,n_j\bigr) \\ &\times \mathrm{Prob}\bigl(n_j=\mathrm{Median}\bigl[\{n_j\}_{j=k-M}^{k+M}\bigr] \bigr).    
\end{split}
\end{align}
Next, we would like to show that the covariance~\eqref{Eq:Cov-nk-med} is equal to ${\sigma^2}/{N_w}$. Indeed, recall that for $N_w$ being odd, the probability that $n_j$ equals to the median is $1/N_w$, see Eq.~\eqref{Eq:Prob}. Further, since all $n_j$ are independent, $\mathrm{Cov}\left(n_k,n_j\right)=\sigma^2\delta_{k,j}$. Thus, the covariance between $n_k$ and the median is $\sigma^2/N_w$. Hence, for large window sizes, the filtered data variance~\eqref{Eq:Var-Med-0} becomes
\begin{align}
\label{eq:filtered-var}
\begin{split}
   \tilde{\sigma}^2 &= \sigma^2 + \sigma^2_{\mathrm{median}}  - 2\frac{\sigma^2}{N_w} \\ & \approx \sigma^2\left[1 - \left(\frac{4-\pi}{2N_w}\right)\right].  
\end{split}
  \end{align}
Therefore, when $N_w\gg 1$, we get 
\begin{equation}
    \label{eq:RMSF-var}
    \tilde{\sigma}^2\rightarrow \sigma^2.
\end{equation} 
Hence, this shows that applying the RMSF preserves the sample variance of the original data in the limit of large window sizes.

\section{RMSF induces no correlation in a white noise stream}\label{App:ACF-RMSF}
We have Eq.~\eqref{eq:RMSF-k} for the RMSF~\eqref{Eq:med-sub} applied to a noise stream:
\begin{align}
 \nonumber    \Tilde{n}_{k} = n_{k}-\mathrm{Median}\left[\{n_j\}_{j=k-M}^{k+M}\right]\,,
\end{align}
and then the RMSF definition with additional lag $l$, so the new index is $k+l$,
\begin{align}
\label{eq:RMSF-k-l}
     \Tilde{n}_{k+l}=n_{k+l}-\mathrm{Median}\left[\{n_j\}_{j=k+l-M}^{k+l+M}\right].
\end{align}
The covariance operator is linear in each of its two variables,
\begin{align}
\label{eq:bilinear}
\begin{split}
 \mathrm{Cov}&(X_1+X_2, Y_1+Y_2) \\ & = \mathrm{Cov}(X_1, Y_1) + \mathrm{Cov}(X_1, Y_2)  \\ & + \mathrm{Cov}(X_2, Y_1) + \mathrm{Cov}(X_2, Y_2).    
\end{split}
\end{align}
For ease, let us define the variables:
\begin{align}
    \begin{split}
        X &= \mathrm{Median}\left[\{n_j\}_{j=k-M}^{k+M}\right] \\ 
        Y &= \mathrm{Median}\left[\{n_j\}_{j=k+l-M}^{k+l+M}\right].
    \end{split}\label{Eq:XY}
\end{align}
Applying Eqs.~\eqref{eq:bilinear} to~\eqref{eq:RMSF-k} and~\eqref{eq:RMSF-k-l}, we arrive at 
\begin{align}
\begin{split}
\mathrm{Cov}&\left( \tilde{n}_{k},\tilde{n}_{k+l}\right) \\&= \mathrm{Cov}\left( {n}_{k},{n}_{k+l}\right) -\mathrm{Cov}\bigl( {n}_{k},Y\bigr)\\
    & -\mathrm{Cov}\bigl( {n}_{k+l},X\bigr)+\mathrm{Cov}\bigl(X, Y\bigr).
    \end{split}
    \label{Eq:Cov}
\end{align}
The first term on the right hand side of Eq.~\eqref{Eq:Cov} is zero since the data points $\{n_k\}$ are independent~\cite[Theorem 4.2.10]{Casella}. The second and third terms on the right hand side of~\eqref{Eq:Cov} are equal to $\sigma^2/N_w$, c.f. derivation of Eq.~\eqref{Eq:Cov-nk-med}. As to the fourth term, we apply the Cauchy-Schwarz inequality~\eqref{eq:Cauchy} to random variables $X$ and $Y$ as in Eq.~\eqref{Eq:XY} and from Eq.~\eqref{Eq:Var-Med}, we get the following:
\begin{equation}
\label{eq:std-med}
  \sigma_X , \sigma_Y \sim \left[\frac{\pi\sigma^2}{2N_w}\right]^{1/2}.
\end{equation}
Therefore, from Eqs.~\eqref{eq:std-med} and~\eqref{eq:Cauchy}, we get:
\begin{align}
\begin{split}
\left|\mathrm{Cov}\left(X, Y)\right)\right| \le \frac{\pi \sigma^2}{2N_w}.    
\end{split}
\label{eq:fourth-term}
\end{align}
Combining~\eqref{eq:fourth-term} with the above observations on the first three terms in~\eqref{Eq:Cov}, we complete the proof that 
the covariance between the filtered data $\tilde{n}_k$ and $\tilde{n}_{k+l}$ with lag $l\neq0$, has order $\mathcal{O}(N^{-1}_w)$. 
The same statement holds true for the ACF. Indeed, $\mathrm{Corr}(\tilde{n}_k, \tilde{n}_{k+l}) = \mathrm{Cov}(\tilde{n}_k, \tilde{n}_{k+l})/\tilde{\sigma}^2$. The numerator is $O(N^{-1}_w)$, while the denominator converges to $\sigma^2$, see~\eqref{eq:RMSF-var}. In other words, the filtered data stream remains white in the limit of large windows. 

\newpage 
\bibliography{References.bib}











`








\end{document}